\documentclass[preprint,aps,prl]{revtex4}
\usepackage{epsfig}
\usepackage{float}

\begin{document}

\title{Van der Waals interactions at surfaces by DFT using Wannier functions} 
\author{Pier Luigi Silvestrelli, Karima Benyahia, Sonja Grubisi\^{c}, 
Francesco Ancilotto, and Flavio Toigo}

\affiliation{
Dipartimento di Fisica ``G. Galilei'',
Universit\`a di Padova, via Marzolo 8, I-35131 Padova, Italy
and DEMOCRITOS National Simulation Center, Trieste, Italy}

\date{\today}

\begin{abstract}
The method recently developed to include Van der Waals interactions
in the Density Functional Theory by using the Maximally-Localized 
Wannier functions, is improved and extended to the case
of atoms and fragments weakly bonded
(physisorbed) to metal and semimetal surfaces, thus opening the
way to realistic simulations of surface-physics processes,
where Van der Waals interactions play a key role.
Successful applications to the case of Ar on graphite and on 
the Al(100) surface,
and of the H$_2$ molecule on Al(100) are presented.
\end{abstract}
\maketitle

\vfill \eject

\noindent
\narrowtext
Understanding adsorption processes is essential to 
design and optimize countless material applications, and
to interpret scattering experiments and atomic-force microscopy.
In particular, the adsorption of rare-gas atoms on metal
and semi-metal surfaces is prototypical\cite{Bruch} for physisorption.
The weak binding of physisorbed closed electron-shell atoms or
saturated molecules like H$_2$ is due to an
equilibrium between attractive long-range Van der Waals (VdW) interactions
and short-range Pauli repulsion.
Density Functional Theory (DFT) is a well-established 
computational approach to study 
the structural and electronic properties of 
condensed matter systems from first principles, and, in particular, to 
elucidate complex surface processes such
as adsorptions, catalytic reactions, and diffusive motions. 
Although current density functionals are able to describe quantitatively
several systems at much lower computational cost than other 
first principles methods, they fail to do so\cite{Kohn} for the 
description of VdW interactions, particularly the leading
$R^{-6}$ term originating from correlated instantaneous dipole 
fluctuations.
The key issue is finding an accurate way to include 
VdW effects in DFT without dramatically increasing the computational cost.
We have recently proposed\cite{PRL} a novel approach, based on the use
of Wannier functions, that 
combines the simplicity of a semiempirical formalism\cite{Grimme} with
the accuracy of the first principles approaches (see for instance
ref.\cite{Langreth04}), and appears
to be promising, being simple, efficient, accurate, and
transferable (charge polarization effects are naturally included).
The results of test applications to small molecules and 
bulk graphite were successful\cite{PRL}.
In this paper we describe improvements in our method which allow
to extend it to the
case of the interaction between an atom or a neutral
fragment and a surface.

Crucial to our analysis is the use of the Maximally-Localized 
Wannier function (MLWF) formalism\cite{Wannier}, that allows
the total electronic density to be partitioned, in a chemically 
transparent and unambiguous way, into individual fragment contributions,
even in periodically-repeated systems.
The MLWFs, $\{w_n({\bf r})\}$, are generated by performing a unitary 
transformation in the subspace of the occupied Kohn-Sham orbitals, 
obtained by a standard DFT calculation, so as to minimize the functional
$\Omega$, defined as :

\begin{equation}
\Omega = \sum_n S_n^2 = 
\sum_n \left( \left<w_n|r^2|w_n\right> - 
\left<w_n|{\bf r}|w_n\right>^2 \right)\;.
\label{spread}
\end{equation}
Besides its spread, $S_n$, each MLWF is characterized also by its
Wannier-function center (WFC); for instance, 
if periodic boundary conditions are used with a cubic supercell of side $L$, 
the coordinate $x_n$ of the $n$-th WFC is defined\cite{Wannier} as

\begin{equation}
x_n = -{L\over {2\pi}}{\rm Im}\; {\rm ln}
\left< w_n | e^{-i{2\pi \over L} x} | w_n \right>\;.
\label{rcenter}
\end{equation}
If spin degeneracy is exploited, every MLWF corresponds
to 2 paired electrons.
Suitable codes\cite{Wannier90} are available,
which allow the efficient generation of the MLWFs, by
adopting a proper $k$-point sampling of the Brillouin Zone (BZ),
which is crucial for metallic systems. 

Starting from these MLWFs the leading
$R^{-6}$ VdW correction term can be evaluated\cite{PRL} by making the
reasonable assumption\cite{Resta} of exponential localization of the MLWFs
in real space, so that each of them is supposed to be an H-like,
normalized function, centered around its WFC position, $r_n$, 
with a spread $S_n$:

\begin{equation}
w_n(|{\bf r - \bf r_n}|) = {3^{3/4} \over {\sqrt{\pi}S_n^{3/2}}}
e^{-{\sqrt{3}\over S_n}|{\bf r - \bf r_n}|}\;.
\label{MLWF}
\end{equation}
Then the binding energy of a system composed of two fragments is given by
$E_b=E_0+E_{\rm VdW}$, where $E_0$ is the binding energy obtained 
from a standard DFT calculation, while the VdW correction is 
assumed to have the form:

\begin{equation}
E_{\rm VdW} = -\sum_{n,l} f_{nl}(r_{nl}){C_{6nl}\over {r_{nl}^6}}\;,
\label{EVdw}
\end{equation}
where $r_{nl}$ is the 
distance of the $n$-th WFC, of the first fragment, 
from the $l$-th WFC of the second one, the sum is over all the 
MLFWs of the two fragments, and the $C_{6nl}$ 
coefficients can be calculated directly 
from the basic information (center positions and spreads) given
by the MLFWs. In fact, using the expression proposed by 
Andersson {\it et al.} (see Eq. (10) of ref.\cite{Langreth96})
that describes the long-range interaction between two
separated fragments of matter:

\begin{equation}
C_{6nl}={3 \over {32\pi^{3/2}}}\int_{|{\bf r}|\leq r_c}d{\bf r}
\int_{|{\bf r'}|\leq r_c'}d{\bf r'}
{\sqrt{\rho_n(r)\rho_l(r')} \over {\sqrt{\rho_n(r)}+\sqrt{\rho_l(r')}}}
={3 \over {32\pi^{3/2}}}\int_{|{\bf r}|\leq r_c}d{\bf r}
\int_{|{\bf r'}|\leq r_c'}d{\bf r'}
{{w_n(r)w_l(r')} \over {w_n(r)+w_l(r')}}\;,
\label{integral}
\end{equation}
where $\rho_n(r) = w_n^2(r)$ is the electronic density corresponding
to the $n$-th MLWF, $C_{6nl}$ is given in a.u., and
the $r_c$, $r_c'$ cutoffs have been 
introduced\cite{Langreth96,Ashcroft}
to properly take into account both the limit of
separated fragments and of distant disturbances in an electron gas.
By using the analytic form (see Eq. (\ref{MLWF})) of the MLWFs,
it is straightforward\cite{PRL} to obtain the 
cutoff expressed in terms of the MLWF spread:

\begin{equation}
r_c = S_n\sqrt{3}\left(0.769 + 1/2{\rm ln}(S_n) \right)\;,
\label{cutoff2}
\end{equation}
and to evaluate very efficiently the multidimensional integral
of Eq. (\ref{integral});
when each MLWF describes 2 paired electrons 
the $C_{6nl}$ coefficients must be multiplied by a $\sqrt{2}$ factor\cite{PRL}.
 
In Eq. (\ref{EVdw}) $f_{nl}(r)$ is a damping function to cutoff
the unphysical behavior of the asymptotic VdW correction at small 
fragment separations.
For this we have chosen the form\cite{Wu,Grimme}:

\begin{equation}
f_{nl}(r) = {1 \over {1+exp(-a(r/R_s-1))}}\;,
\label{fr}
\end{equation}
where\cite{Grimme} $a \simeq 20$ (the results are almost
independent on the particular value of this parameter), 
and $R_s = R_{\rm VdW}+R'_{\rm VdW}$ is the sum of the VdW radii 
of the MLWFs. In ref. \cite{PRL} $R_{\rm VdW}$ was determined as 
the radius of the 0.01 Bohr$^{-3}$ electron density contour.
For the present applications to extended systems with metal
or semimetal surfaces, after extensive testing, we found that 
a better choice is to
equate $R_{\rm VdW}$ to the cutoff radius introduced
in Eqs. (5) and (6), $R_{\rm VdW}=r_c$, which has the 
additional advantage of not being 
dependent on any given electron density threshold.
It should be stressed that the results reported in ref. \cite{PRL}, relative
to isolated fragments and bulk graphite, are essentially unchanged 
if recomputed by adopting this new $R_{\rm VdW}$ definition. 

The $E_0$ binding energy can be obtained 
from a standard DFT calculation (for instance, using the
Quantum-ESPRESSO\cite{ESPRESSO} ab initio package),
with the Generalized Gradient Approximation in the
revPBE flavor\cite{revPBE}. This choice\cite{Grimme,Langreth04} is motivated 
by the fact that revPBE is fitted to the exact Hartree-Fock exchange,
so that the VdW correlation energy only comes
from the VdW correction term, as described above, without any
double-counting effect.
The evaluation of the VdW correction as a post-DFT
perturbation, using the revPBE electronic density distribution,
represents an approximation because, in principle, a full
self-consistent calculations should be performed; however
recent investigations\cite{Langreth07} on different systems 
have shown that the effects due to the lack of self-consistency
in VdW-corrected DFT schemes are negligible.
The method described above can be refined by
considering the anisotropy\cite{PRL} of the MLWFs
(details will be published
elsewhere\cite{ELSE}); however, since previous calculations\cite{PRL} 
have shown that this has small effects, it has been not included in the
present applications.
Remarkably, the whole procedure of generating the MLWFs and evaluating the
VdW corrections represents a negligible additional computational cost,
compared to that of a standard DFT calculation. 

We have applied our method to the case of Ar on graphite and on
the Al(100) surface, and of the saturated H$_2$ molecule on Al(100).
Adsorption on graphite has been modeled using an hexagonal
supercell containing 36 C atom distributed over 2 graphene sheets,
with a sampling of the BZ limited to the $\Gamma$ point
(preliminary tests and previous studies\cite{Bichoutskaia,Tkatchenko}
show that these choices are adequate); in the case of the 
Al(100) surface the supercell was orthorhombic with a surface slab made of 
32 Al atoms distributed over 4 layers, 
and a $2\times2\times1$ sampling of the BZ
was used; no appreciable difference in the equilibrium
properties was observed in test calculations with a thicker slab of 64 Al atoms
over 8 layers (of course a thicker slab would be instead necessary to
describe well the far-from-the-surface, asymptotic behavior, where the
binding energy is expected to decay as $z^{-3}$, $z$ being the 
fragment-surface distance). 
For a better accuracy,
in these applications it has been necessary to modify our algorithm 
in such a way to include interactions of
the MLWFs of the physisorbed fragments (Ar or H$_2$) not only with
the MLWFs of the underlying surface, within the reference supercell, 
but also with a sufficient
number of periodically-repeated surface MLWFs (in any case, given the 
$R^{-6}$ decay of the VdW interactions the convergence is rapidly achieved).
Electron-ion interactions were described using norm-conserving
pseudopotentials (in the case of Al only the 3 valence electrons per atom
were explicitly included).
In principle, for evaluating adsorption properties in periodically-repeated,
asymmetric configurations, one should add a dipole correction\cite{Bengtsson}
that compensates for the artificial dipole field introduced by the
periodic boundary conditions; however we have checked that, in our 
cases, this correction is very small (just a few meV in the binding 
energy of Ar on Al(100)).

The absorption of noble gases on graphite and on metal surfaces
has been studied extensively 
over the years\cite{Bichoutskaia,Diehl} because it serves as the
paradigm of weak adsorption.
Actually, despite the conceived ``simplicity'' of these 
systems, even the most basic question (what is the preferred 
adsorption site ?) has not been answered in an entirely satisfactorily way. 
In principle, due to the non-directional character of the VdW
interactions, sites that maximize the
coordination of the adsorbate atom were expected,
so that it was typically assumed that
the adsorbate occupies the maximally coordinated {\it hollow} site.
The actual scenario is more complex:
for Xe and Kr a clear preference is found\cite{Diehl,Dasilva} for 
adsorption on metallic surfaces in the low-coordination {\it top} sites
(this behavior was attributed to the delocalization
of charge density that increases the repulsive effect at the
{\it hollow} sites relative to the {\it top} site and lifts the potential well
upwards both in energy and height); for Ar the situation seems 
to be different:
comparison of theoretical and experimental results\cite{Diehl}
would suggest that the {\it hollow} sites are favored for Ar on Ag(111)
and on graphite, although, in this latter case, this 
configuration is preferred over two other possible sites ({\it top} 
and {\it bridge}), by only a few meV\cite{Bichoutskaia}.

Ar on Al(100) represents a critical test for our method, in fact
the Al case is particularly challenging for a Wannier-based
scheme since Al is the metal which most closely approximates
a free electron gas system: hence the electronic charge is relatively 
delocalized and the assumption of exponential localization of the MLWFs
is no longer strictly valid\cite{Resta}.
However the following results show that, even in this case, our method
works and this does not come to a surprise.
In fact, on the one hand, the MLWF technique has been efficiently 
generalized also to metals\cite{Souza,Iannuzzi}, on the other,
bonding in metallic clusters and in fcc bulk metals (like Al) can be
described in terms of H-like orbitals localized on tetrahedral 
interstitial sites\cite{Souza}, which is just in line
with the spirit of the present scheme.

In the Tables I and II we report our computed binding energies and
equilibrium fragment-surface distances, compared to the most
reliable (to our knowledge) experimental and 
theoretical reference data, and to the results of LDA calculations
(we have also reported the values obtained in ref.\cite{PRL} for
Ar interacting with the benzene molecule).
As can be seen, the general performance of the method is quite
satisfactory; in fact, the improvement achieved by including the VdW
correction, with respect to the pure revPBE scheme 
(which gives completely unphysical results, namely a potential well 
very small and located too far from the surface) is dramatic.

In the case of Ar on graphite, the {\it hollow} configuration 
is energetically favored,
although by just a few of meVs with respect to the other two
configurations, in good agreement with previous studies\cite{Bichoutskaia}.
Interestingly, our estimated Ar-graphite surface distance essentially
coincides with the sum of the Ar and C VdW literature radii 
(1.88+1.73=3.61 \AA),
a behavior experimentally observed in the related case of Xe adsorbed 
on graphite\cite{Diehl}.
Moreover, the fact that the Ar-graphite distance is not appreciably
smaller in the {\it hollow} site, compared to the {\it top} one, could
be rationalized in terms of the potential lifting, due to increased repulsion,
mentioned above.  
Note that the binding energy of Ar on graphite is considerably larger
than that of the Ar-benzene complex, although the equilibrium distance
is similar; this behavior is clearly due to the VdW interaction of Ar with
the electronic charge outside the underlying C ring, and is not 
reproduced by the LDA approach which favors short-range interactions. 

Concerning Ar on Al(100) (see also Fig. 1), specific experimental values are not
available, however the experimental binding energy of Ar
on several other metals is found to be in the range between 30 and  
100 meV\cite{Diehl,Kirchner,Unguris}
(between 70 and 85 meV\cite{ChengPRB} for noble metals), in 
agreement with our VdW-corrected results. 
We also mention old
theoretical estimates of a binding energy of about 200 meV 
\cite{Andriotis}, and of 70 meV using a jellium model\cite{Lang}.

In the case of H$_2$ on Al(100) 
the molecule is essentially a free rotor in the physisorption 
regime\cite{Chizmeshya} and its interaction with the substrate
exhibits only a slight anisotropy; moreover the effect of
changing the position of the molecule with respect to the substrate
is small, so that we report only the results relative to a single,
representative configuration. Even for this extremely weakly bonded
system the results are in good agreement with the reference values
(we also mention that the binding energy of H$_2$ on Mg is
is predicted to be 17 meV\cite{Chizmeshya}).
  
Looking at the tables, on can see that the binding energies
are reasonable reproduced by the LDA scheme,
although this is actually accidental (the well-known
LDA overbinding, due to the overestimate of the long-range part of the 
exchange contribution, somehow mimics the missing VdW interactions), moreover
the equilibrium distances are clearly underestimated. 

In conclusion, we have extended our recently developed scheme,
to include VdW interactions
in the DFT by using the MLWFs, to the case of fragments weakly bonded
(physisorbed) to metal and semimetal surfaces, and we have reported
results of applications to the case of Ar on graphite and on
the Al(100) surface,
and of the H$_2$ molecule on Al(100). 
The good performances of the method clearly indicate that it can be
very useful to investigate many realistic surface-physics processes,
where VdW interactions play a key role. 
Of a particular value is the possibility of dealing with metal surfaces
(insulating surfaces could be somehow treated even using atom-based 
semiempirical approaches\cite{Grimme}). 
Finally it must be stressed that a large area for future improvements 
of the method exists. In fact, different, more
sophisticated schemes to utilize the MLWFs could be developed:
for instance, one could adopt {\it{gaussians}} instead of exponential,
H-like, functions, because multidimensional integrals are more
easily evaluated; orbitals of symmetry different from the $s$-like one
could be used for specific applications; partially occupied 
MLWFs\cite{Thygesen}, with improved localization and symmetry properties,
could be introduced (particularly for metallic
systems); different damping functions, and improved, reference DFT functionals, 
with respect to revPBE, could be chosen,...

We acknowledge 
allocation of computer resources from INFM
``Progetto Calcolo Parallelo'' and the support of Padova University
through project CPDA077281-07.

\vfill
\eject

\begin{table}
\caption{Binding energy, in meV, of Ar on graphite and Al(100) and
H$_2$ on Al(100), 
computed using the standard DFT-revPBE
calculation, and including the VdW correction, 
compared to the LDA result, and  available theoretical and
experimental (in parenthesis) reference data.}
\begin{center}
\begin{tabular}{|c|c|c|c|c|}
\hline
system & revPBE & revPBE+VdW & LDA & ref. \\ \tableline
\hline
Ar-gr. {\it hollow}         & -2 &-133 & -76 & -116$^a$, -111$^b$ (-119$^c$)\\ 
Ar-gr. {\it top}            & -2 &-131 & -69 &           -106$^b$\\     
Ar-gr. {\it bridge}         & -2 &-128 & -69 &           -107$^b$\\
Ar-benzene            & -2 & -66 & -72 &  -65$^d$ (-49$^e$)\\
\hline
Ar-Al(100) {\it hollow}     & -3 & -72 & -71 & ---  \\
Ar-Al(100) {\it top}        & -3 & -71 & -66 & ---  \\
\hline
H$_2$-Al(100)         & -2 & -20 & -24 & -19$^f$ (-28$^g$) \\ 
\hline
\end{tabular}                                                
\tablenotetext[1]{Reference\cite{Tkatchenko}.}
\tablenotetext[2]{Reference\cite{Bichoutskaia}.}
\tablenotetext[3]{Reference\cite{Unguris}.}
\tablenotetext[4]{Reference\cite{Lilienfeld}.}
\tablenotetext[5]{Reference\cite{Jin}.}
\tablenotetext[6]{Reference\cite{Cheng}.}
\tablenotetext[7]{Reference\cite{Andersson}.}
\end{center}
\label{table1}                                  
\end{table}

\begin{table}
\caption{Equilibrium distance, in \AA, of Ar on graphite and Al(100) and
H$_2$ on Al(100), 
computed using the standard DFT-revPBE
calculation, and including the VdW correction, 
compared to the LDA result, and  available theoretical and
experimental (in parenthesis) reference data.}
\begin{center}
\begin{tabular}{|c|c|c|c|c|}
\hline
system & revPBE & revPBE+VdW & LDA & ref. \\ \tableline
\hline
Ar-gr. {\it hollow}         & 4.69 & 3.61 & 3.16 & 3.33$^a$, 3.32$^b$ \\
Ar-gr. {\it top}            & 4.95 & 3.61 & 3.22 &           3.37$^b$ \\
Ar-gr. {\it bridge}         & 4.69 & 3.55 & 3.22 &           3.37$^b$ \\
Ar-benzene            & 4.79 & 3.57 & 3.27 & 3.41$^c$ (3.68$^d$)\\
\hline
Ar-Al(100) {\it hollow}     & 5.34 & 4.66 & 3.48 & ---  \\
Ar-Al(100) {\it top}        & 5.34 & 4.88 & 3.57 & ---  \\
\hline
H$_2$-Al(100)         & 5.08 & 4.62 & 3.23 & --- \\ 
\hline
\end{tabular}                                                
\tablenotetext[1]{Reference\cite{Tkatchenko}.}
\tablenotetext[2]{Reference\cite{Bichoutskaia}.}
\tablenotetext[3]{Reference\cite{Lilienfeld}.}
\tablenotetext[4]{Reference\cite{Jin}.}
\end{center}
\label{table2}                                  
\end{table}

\vfill
\eject

\pagestyle{empty}
\begin{figure}
{\vskip 1.3cm}
\centerline{
\includegraphics[width=17cm]{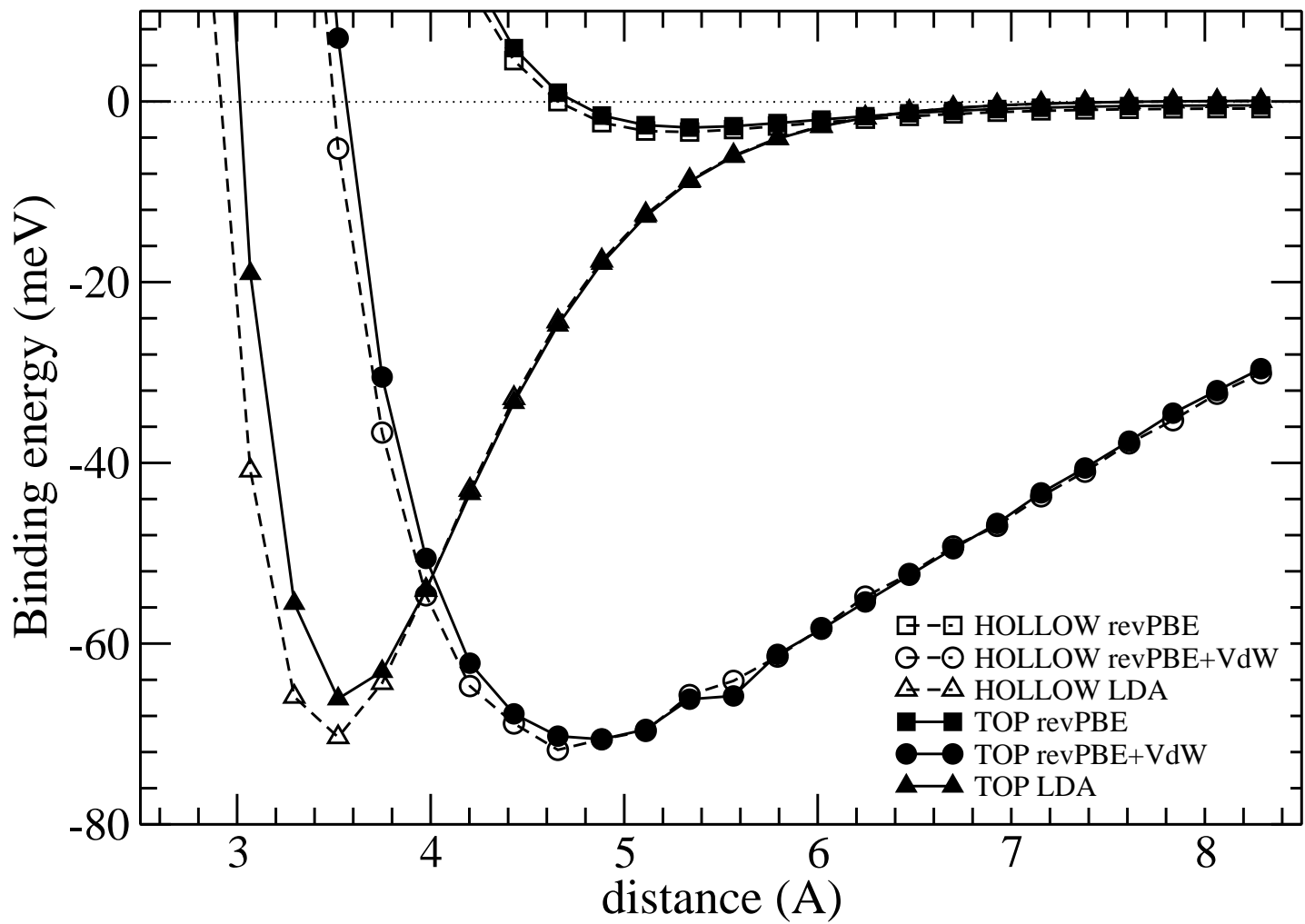}
}
\caption{Binding energy of Ar on Al(100) as a function of the
distance from the surface.}
\label{fig1}
\huge
\end{figure}

\end{document}